\documentclass{article}
\usepackage{amsmath}    
\usepackage{multirow}   
\usepackage{amssymb}    
\usepackage{graphicx}   
\usepackage{subfigure}  
\usepackage{setspace}   
\usepackage{booktabs}   
\usepackage{geometry}   
\usepackage{booktabs}   
\usepackage[numbers,compress]{natbib}   
\usepackage{authblk}    
\usepackage{textcomp}   
\usepackage{makecell}
\usepackage{color}
\renewcommand{\thesubfigure}{(\alph{subfigure})}
\makeatletter
\renewcommand{\p@subfigure}{}
\renewcommand{\@thesubfigure}{\thesubfigure\hskip\subfiglabelskip}
\makeatother
\def \AA{\text{\r{A}}}
\title{First principles study on the mechanism of abnormal viscosity change of pure Al \& Pb melts based on Wulff cluster model}
\date{}                 
\author[a]{Anchen Shao}
\author[a]{Lina Hu}
\author[b]{Lin Song}
\author[a]{Minghao Hua}
\author[a]{Jiajia Xue}
\author[a]{Shuang Wu}
\author[a]{Xuelei Tian \thanks{Corresponding Author: Tel.: +86 531 88392319; E-mail: tianxuelei@sdu.edu.cn}} 
\author[a]{Xiaohang Lin \thanks{Corresponding Author: Tel.: +86 531 88392727; E-mail: lxh12345@sdu.edu.cn}}
\affil[a]{Key Laboratory for Liquid-Solid Structural Evolution and Processing of Materials, Ministry of Education, Shandong University, Jinan 250061, People's Republic of China}
\affil[b]{Shandong Laboratory of Yantai Advanced Materials and Green Manufacture, Yantai, Shandong province, 264000, China}

\begin{document}
\maketitle
\begin{abstract}
  In this paper, the Wulff cluster model combined phonon calculation is used to investigate the relationship between the structure of metallic melts (Pb/Al) and the abnormal viscosity change. 
  Although absolute value of the surface energy does not change significantly with temperature, the Wulff shape changed evidently.
  When temperature raise to 975K, Pb(321) surface that has the highest interaction strength completely disappears, while the abnormal viscosity drop happens at the same temperature range.  
  Oppositely, Al(100) surface that has the lowest interaction strength disappears when temperature is higher than 1075K, and at the same temperature range, the abnormal viscosity rise has been observed.
  The abnormal viscosity drop corresponds to the disappearance of surface with the highest interaction strength (Pb(321) surface), while the rise one corresponds to the disappearance of surface 
  with the lowest adsorption energy (Al(100) surface).  
  All evidence indicates that the uniformity between abnormal viscosity change and the change of Wulff shape is not a coincidence. 
  A possible mechanism of the abnormal viscosity change is that it caused by the significant structure change of cluster (short range ordering) in the metallic melts.
   \\ \hspace*{\fill} 
   \\
   \noindent{\textbf{Keywords: } Pb/Al melt, abnormal viscosity change, Wulff cluster model, DFT, phonon calculation}
\end{abstract}

\section{Introduction}
The viscosity is an important physical parameters of metal melts, which attract much attention \cite{rev01,rev02,rev03}. Most traditional hot working processes, especially casting, have been strongly influenced by
the viscosity of metal melts. For example, the process defects are often caused by metal viscosity control problems \cite{VisManu01,VisManu02,VisWelding,VisWelding02} . 
The atomic scale understanding of melt viscosity
is helpful to better design the process parameters of various hot working processes and material preparation processes. 
Unfortunately, the micro-image of melt viscosity is still unclear.

It is generally believed that the viscosity change of high-temperature melt is linearly related to the atomic activation energy, that is, the change characteristics meet the Arrhenius formula\cite{Anu,Anu01,RevAll},
Most experiments are basically consistent with this conclusion, but many metals have some special phenomena that lead to the breaking point of the temperature-viscosity curve at a specific temperature
\cite{AlVis01,GaNeutron,moleRubiCollec,PbSnAnu,PbVis01,RubiVolVis,RuNeutron,VisClusElec},
As the temperature increases, some of the metals (like Pb) which was found in the temperature-viscosity curves will appear an abnormal viscosity drop compared to the prediction value of Arrhenius formula\cite{ChuweiVis,HouPbVis}, 
others (like Al) will 
show an abnormal viscosity rise than prediction value\cite{AlVis01,ChuweiVis}.

Researchers have explained these changes from different perspectives, such as coordination number changes, density change, 
coordination layer spacing change and cluster structure transition\cite{moleRubiCollec,RubiVolVis,ChuweiVis,Cornum,Dens}. Although these reasons can explain the phenomena of the abnormal viscosity change 
to some extent,
the mechanism under atomic level hardly be understood.

In order to investigate the abnormal viscosity change under atomic level, a reasonable model which could accurately describe the microstructure of metallic melts is needed.
In previous researches, we proposed a model, so called Wulff cluster model, which uses the most probable cluster structure 
to describe the short-range ordering structure distribution in metallic melts under the thermal equilibrium state.  
We assume that the short-range ordering structure is determined by metal Wulff construction with the crystal structures inside, 
the size is given by the PDF of the experimental XRD results of metallic melts.  
The Wulff shape is obtained by surface energies in different crystal directions by the following way.
Draw planes along different crystal directions to ensure that the distance from the plane to the origin is in a fixed proportion to the lowest surface energy of the crystal plane family 
corresponding to the crystal direction. After all planes are drawn, the minimum 
space containing the origin surrounded by all planes is the Wulff shape of the crystal.
After experimental 
verification, the structure of the cluster actually accurately 
reflects the statistical average of the cluster structure characteristics in the metal melt.
Our previous studies show that the Wulff cluster model is a reasonable model to describe pure metals, biphasic homogeneous alloys, eutectic alloys, metal-nonmetal eutectic alloys 
and intermetallic compound melts systems. 
The calculated XRD results of the model are in good agreement with the results of high temperature X-ray diffraction (HTXRD) experiments \cite{FES01,FES02,FES03,FES04}. 
The structure of the cluster actually accurately reflects the statistical average of the cluster structure characteristics in the metal melt, 
which show that the Wulff cluster model is a reasonable model to describe pure metals, biphasic homogeneous alloys, eutectic alloys and intermetallic compound melts systems. 
In this case, it is reasonable to use the Wulff cluster model as the feature structure in order to investigate the viscosity of the metallic melts.

But temperature effect has been ignored in the Wulff cluster model. Some studies have pointed out that after considering the change of phonon free energy with temperature, the Gibbs free energy 
of each orientation of crystal plane will change obviously, which will 
lead to the change of Wulff shape.

Generally speaking, the metallic melts can be divided into two part: cluster and free atoms. Obviously free atoms have significantly different binding capacities on different surfaces. 
So it is reasonable to speculate that the shape change of Wulff cluster model considering the temperature effect may be related 
to the breaking point of temperature-viscosity curve of the metallic melts.
As two widely used metals, Al and Pb have breaking points in their melt temperature-viscosity curves, 
which have been mentioned in previous studies\cite{AlVis01,PbSnAnu,PbVis01,HouPbVis}. 
In this paper, starting from the influence of phonon free energy on interface energy, we study the change of Wulff shape of metal Al and Pb with respect to the temperature, 
and put forward a new explanation for the sudden change of melt viscosity with temperature.

\section{Calculation Methods}

In this work, Vienna Ab initio Simulation Package 6.0(VASP6.0) is used for DFT calculation, 
within the generalized gradient (GGA) approximation to describe the exchange-correlation effects, employing the Perdew-Burke-Ernzerhof (PBE) 
exchange-correlation functional.\cite{PBE01,PBE02}, The ionic cores are represented by projector augmented wave (PAW)
potentials as constructed by Kresse and Joubert\cite{PAW01,PAW02}.

The structural optimization and cut-off energy are selected as Pb(550eV) and Al(600eV). For the surface energy calculations, the Monkhorst–pack method and the division steps of $11\times 11 \times 1$ 
are used to process the K point in the reciprocal space. 
The relaxation termination condition is that the interatomic interaction force is less than $0.015 {\rm eV}\cdot \AA^{-1}$.
In previous studies, it has been proved that the relative proportion of solid-liquid interface energy and surface energy under the approximation of equivalent solute 
model is very close \cite{FES01, vaspsol01, vaspsol02}. Therefore, surface energy is directly used to construct Wulff shape in this study.

The double-sided slab model is used to calculate the surface energy of different crystal plane families, that is, remove specific atoms from the perfect supercell to form a vacuum 
layer and expose the required crystal plane. The lattice basis vector $\mathbf{C}$ of the supercell is adjusted to be perpendicular to the exposed surface and the other two basis vectors.

Some atoms of the formed atomic layer away from the two surfaces along the normal direction are fixed, while other atoms close to the surface and 
constituting the surface are subject to relaxation calculation until the relaxation termination conditions are met. The thickness of vacuum layer is set to $15\AA$ to 
avoid the calculation error caused by the interaction between surface atoms. After the structure is optimized, the system energy at this time is taken as $E_ {DFT}{ }$.

To consider temperature effect, the Gibbs free energy including vibrational entropy by phonon calculation is used in this paper.
In phonon calculation, for bulk simulation, a  $2\times 2 \times 2$ expanded supercell is used with the k-points of $7\times 7 \times 7$ mesh to calculate the Hessian matrix. 
while the k-points of $5\times 5 \times 1$ mesh are used for the double-sided slab model.
Phonon dispersion relations and phonon density of state (PDOS) can be obtained by Hessian matrices, then partition function will become calculable, and Gibbs free energies of phonon 
are calculated. 
The software Phonopy is used to calculate the phonon Gibbs free energy at different temperatures as $E_ {phono}{ }$.

The Gibbs free energy of system is given by:

\begin{align}
  E_{Gibbs}=E_{DFT}+E_{phono} \label{etot}
\end{align}

The surface energy(with temperature) $\gamma$is given by the following equation:

\begin{align}
  \gamma=\frac{1}{2A}(E_{Gibbs.slab}-NE_{Gibbs.bulk}) \label{surfef}
\end{align}

In the above formula, $A$ is the area exposed by the crystal plane in the supercell, which can be given by the supercell base vector 
product $\mathbf{a} \times \mathbf{b}$ ($E_{Gibbs.slab}$ and $E_ {Gibbs.bulk}$ are the total energy $E_{Gibbs}$ of the calculated system),
and $N$ is the number of cells contained in the supercell.

\section{Results and Discussion}
 
According to the Wulff cluster model, the most probable structure of short-range ordering in metallic melts is determined by Wulff construction with the crystal structures inside\cite{FES01,FES02,FES03,FES04}. 
Obviously, the inside structures would not change with temperature significantly, 
but what about the shape of the cluster? Associating the viscosity change mentioned above, is this special phenomenon related to the shape change? In the following part, pure metal Aluminum and Lead are investigated separately to answer these questions.

\subsection{The Wulff shape in Pb \& Al melts}

As the first step of the investigation, the WUlff shape without temperature effect should be determined. 
To cleave slab models, the lattice parameters should be determined as the first step. 
The value of Al (FCC) and Pb (FCC) are 4.04 Å and 5.02 Å, which is only about 1\% different from the experimental data\cite{WebEle}.

\begin{table}[htbp]
  \centering
  \caption{DFT Calculation results of surface energies of Al and Pb}
  \label{tab:SurfEne}
  \begin{tabular}{p{3cm}<{\centering}p{3cm}<{\centering}p{3cm}<{\centering}p{3cm}<{\centering}}
     \toprule
     Al & $\gamma/{\rm 10^{-2}eV\cdot\AA^{-2}}$ & Pb & $\gamma/{\rm 10^{-2}eV\cdot\AA^{-2}}$ \\ 
     \midrule
     Al(100) & $5.78$ & Pb(100) &$2.06$\\
     Al(110) & $6.19$ & Pb(110) &$2.11$\\
     Al(111) & $5.09$ & Pb(111) &$1.62$\\
     Al(211) & $5.81$ & Pb(211) &$1.99$\\
     Al(210) & $6.29$ & Pb(210) &$2.04$\\
     Al(221) & $5.93$ & Pb(221) &$1.89$\\
     Al(321) & $6.12$ & Pb(321) &$2.13$\\
     \bottomrule
  \end{tabular}
\end{table}

\begin{figure}[htbp]
  \centering
  \subfigure[]{\includegraphics[scale=0.075]{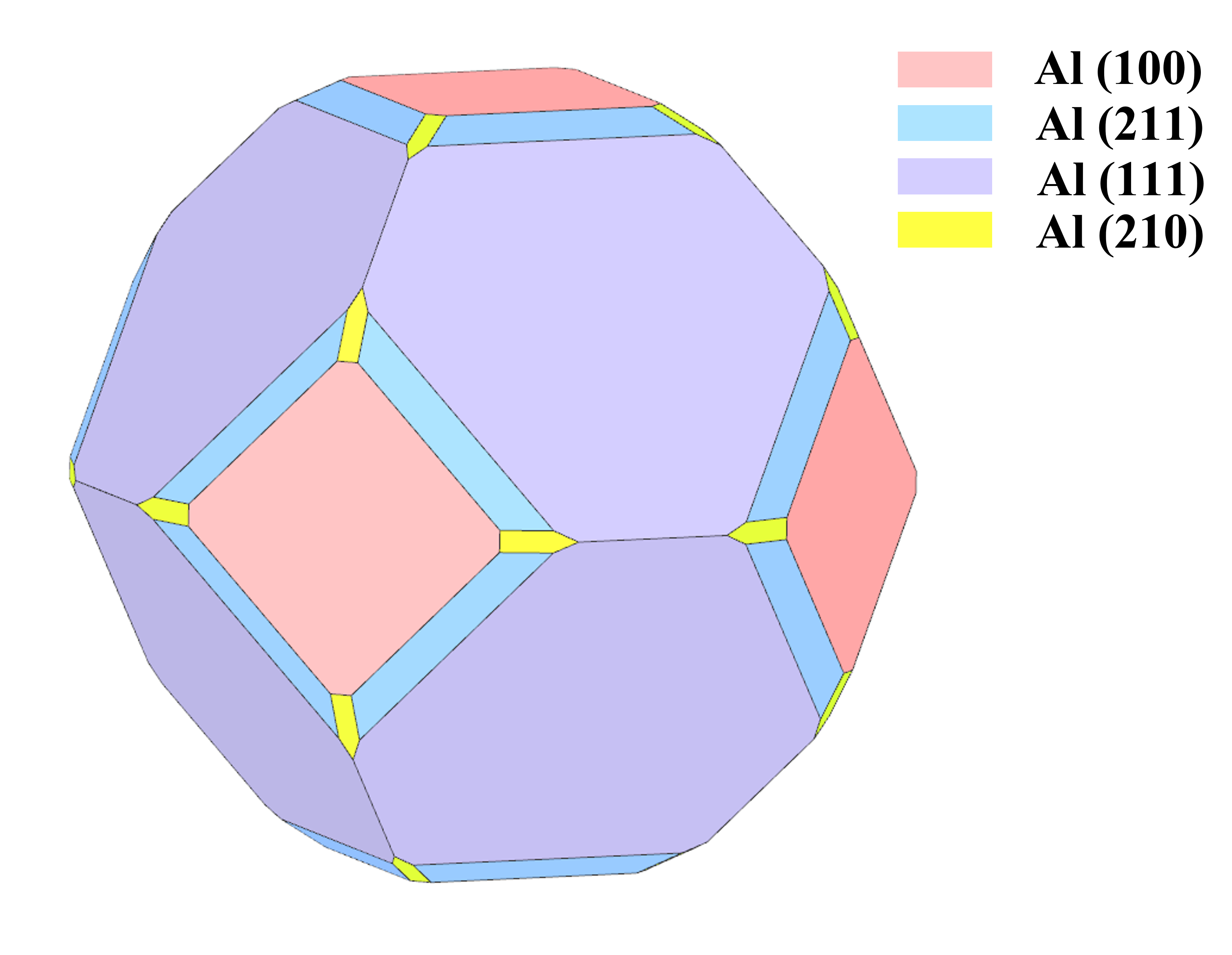} \label{subfig:aldftw}} \qquad 
  \subfigure[]{\includegraphics[scale=0.075]{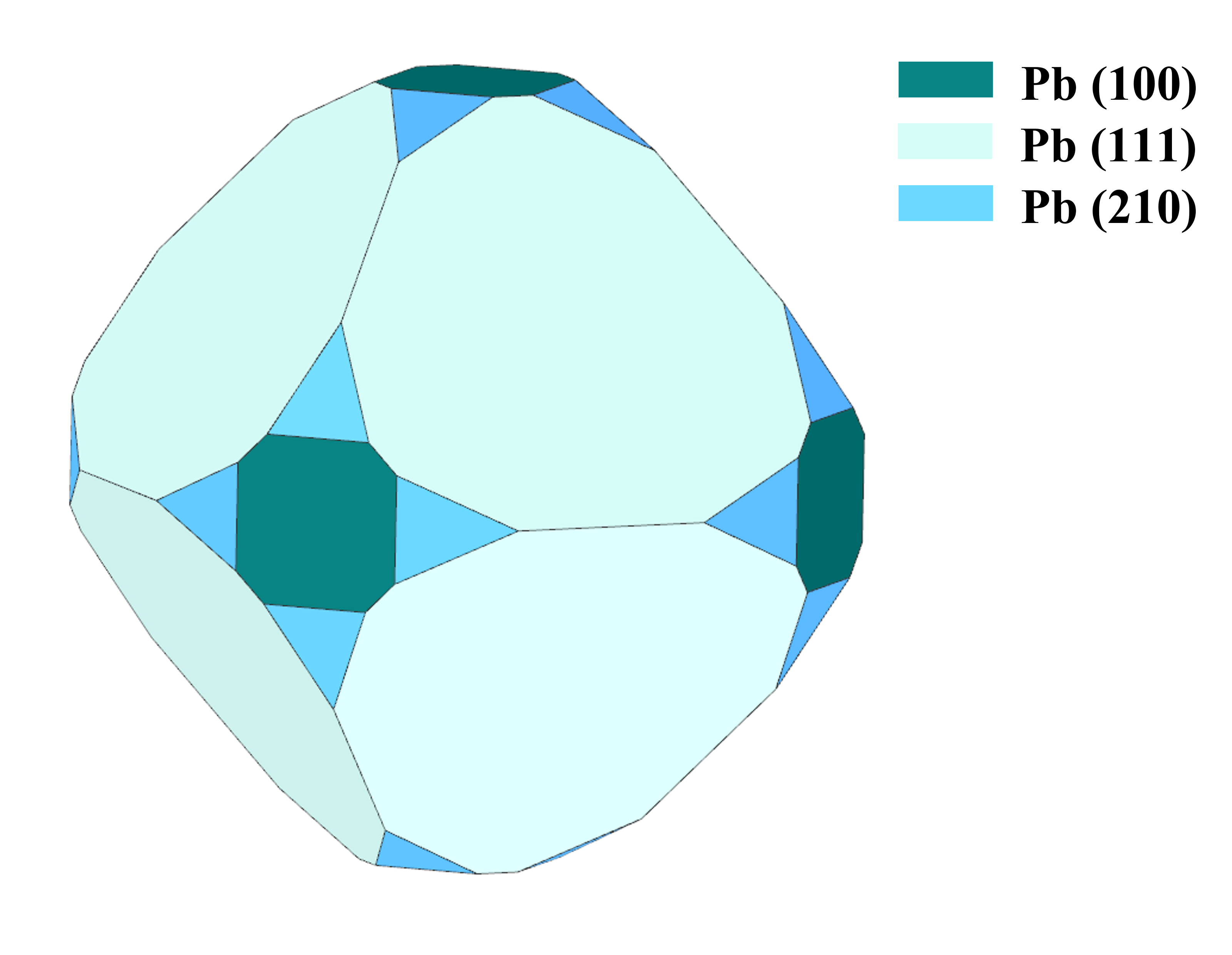} \label{subfig:pbdftw}} \qquad
  \caption{Wulff shape of \subref{subfig:aldftw}Al and \subref{subfig:pbdftw}Pb}
  \label{fig:dftwulff}
\end{figure}

The surface energy of each surface without the phonon energy and the Wulff shape is shown in the Tab.\ref{tab:SurfEne} and Fig.\ref{fig:dftwulff}.
Lead has a face centered cubic lattice. 
The Pb(111) surface is the most stable surface that accounting for 81.4\% of the total surface area. The surface energy of the Pb(221) and 
Pb(211) surfaces with obvious stepped arrangement follows closely, but for geometrical reason those two surfaces disappear from Wulff shape,
for the same reason,Pb(100) and Pb(210) appear even they have a relatively high surface energy.
And the Pb(110) surface has the highest energy and totally disappears. 
This makes the Wulff shape of Pb as an octahedron as a whole(Figure.\ref{fig:dftwulff}\subref{subfig:pbdftw}).

The results of aluminum are similar as that of lead. Al(111) is still the most stable surface, accounting for 66.4\% 
of the total surface area, Al(211) appears but only accounting for 12.8\%,and still reflects the octahedral appearance(Figure.\ref{fig:dftwulff}\subref{subfig:aldftw}).

Obviously, without temperature effect, the structures of the short-range ordering do not change except the cluster size. If Gibbs free energy is used instead of surface energy, 
in other words the temperature effect has token into account, it will be totally different story.


\subsubsection{The lead clusters with temperature effect}

\begin{figure}[htbp]
  \centering
  \includegraphics[scale=0.4]{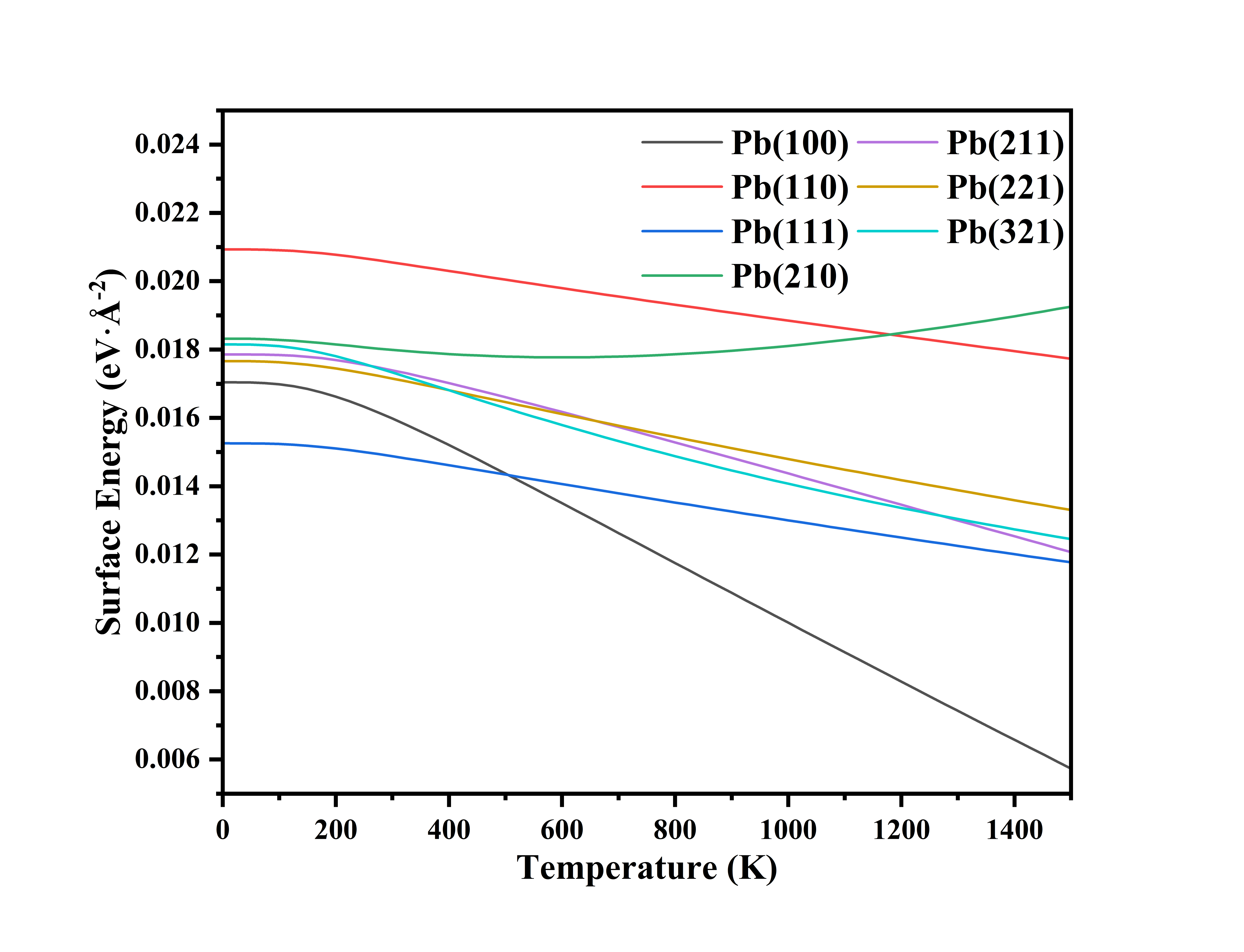}
  \caption{Pb surface energy of different planes with temperature considering phonon Gibbs free energy}
  \label{fig:PbSurfE}
\end{figure}

The surface energy(with temperature)-temperature curve is shown in Figure.\ref{fig:PbSurfE}.
Obviously, most surface energies decrease with inscreasing temperature, except Pb(210) which have a extremum point at about 600K and then the surface energy 
become positive correlation with temperature. Noted that the surface energy(with temperature) of Pb(100) shows a faster decreasing speed than others, 
At about 450K, Pb(100) becomes the most stable surface, which is even more stable than Pb(111) surface. 
This may indicates that Pb(100) will become far more stable in high temperature.

\begin{figure}[h]
  \centering
  \includegraphics[scale=0.175]{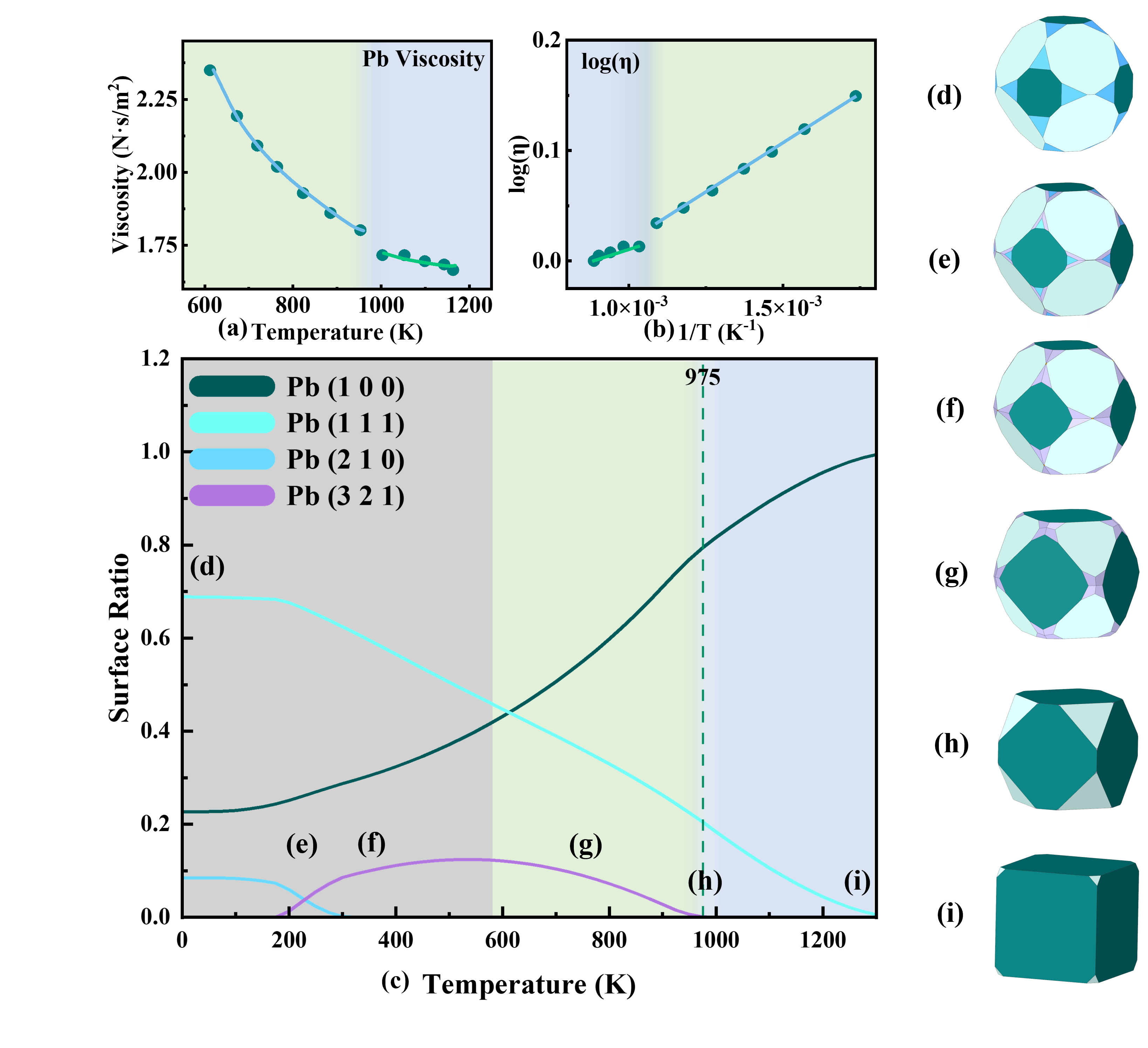}
  \caption{(a)Viscosity-temperature curve of Pb melts, (b)viscosity logarithm-temperature reciprocal curve of Pb melts\cite{HouPbVis}, (c)surface ratio-temperature curve of Pb Wulff shape,
  corresponding Wulff shape are demonstrated in (d)0K (e)250K (f)350K (g)750K (h)975K (i)2500K.}
  \label{fig:PbPho}
\end{figure}

As mentioned above, the viscosity-temperature curve of Pb meets the Arrhenius formula in the range of green (lower than 930K) and blue (higher than 1020K) shown in fig.\ref{fig:PbPho}a.
There is an obvious viscosity breaking point around 975K , as well as viscosity logarithm-temperature reciprocal curve shown in fig.\ref{fig:PbPho}b.
Fig.\ref{fig:PbPho}c displays the change of the surface ratio in Wulff shape with temperature. Surface ratio is defined as the ratio of a group of certain surface area to the total area.
Although absolute value of the surface energy does not change significantly with temperature, the Wulff shape changed evidently. 
Pb(210) surface has been replaced by Pb(321) at a temperature about 275K(fig.\ref{fig:PbPho}e), which absolutely happens in the solid state (area with grey background). 
Then at the temperature a little higher than melting point, the occupation ratio of Pb(100) exceed that of the Pb(111).
In the range of area with green background color, the Wulff shape become more likely a hexahedron rather than a octahedron (fig.\ref{fig:PbPho}g).
At about 975K, Pb(321) surface completely disappears, which is almost the same as the viscosity leap point(marked with a fuzzy boundary between green and blue background colors in fig.\ref{fig:PbPho}a b c),
this indicates that there may be some correlation between the two.
Then at 1300K, the Wulff shape will become totally a hexahedron and Pb(111) disappears. The relationship between the Wulff shape and the  viscosity change will be discussed in section \ref{sec:wulffVis}.

\subsubsection{The Aluminum clusters with temperature effect}

\begin{figure}[htbp]
  \centering
  \includegraphics[scale=0.4]{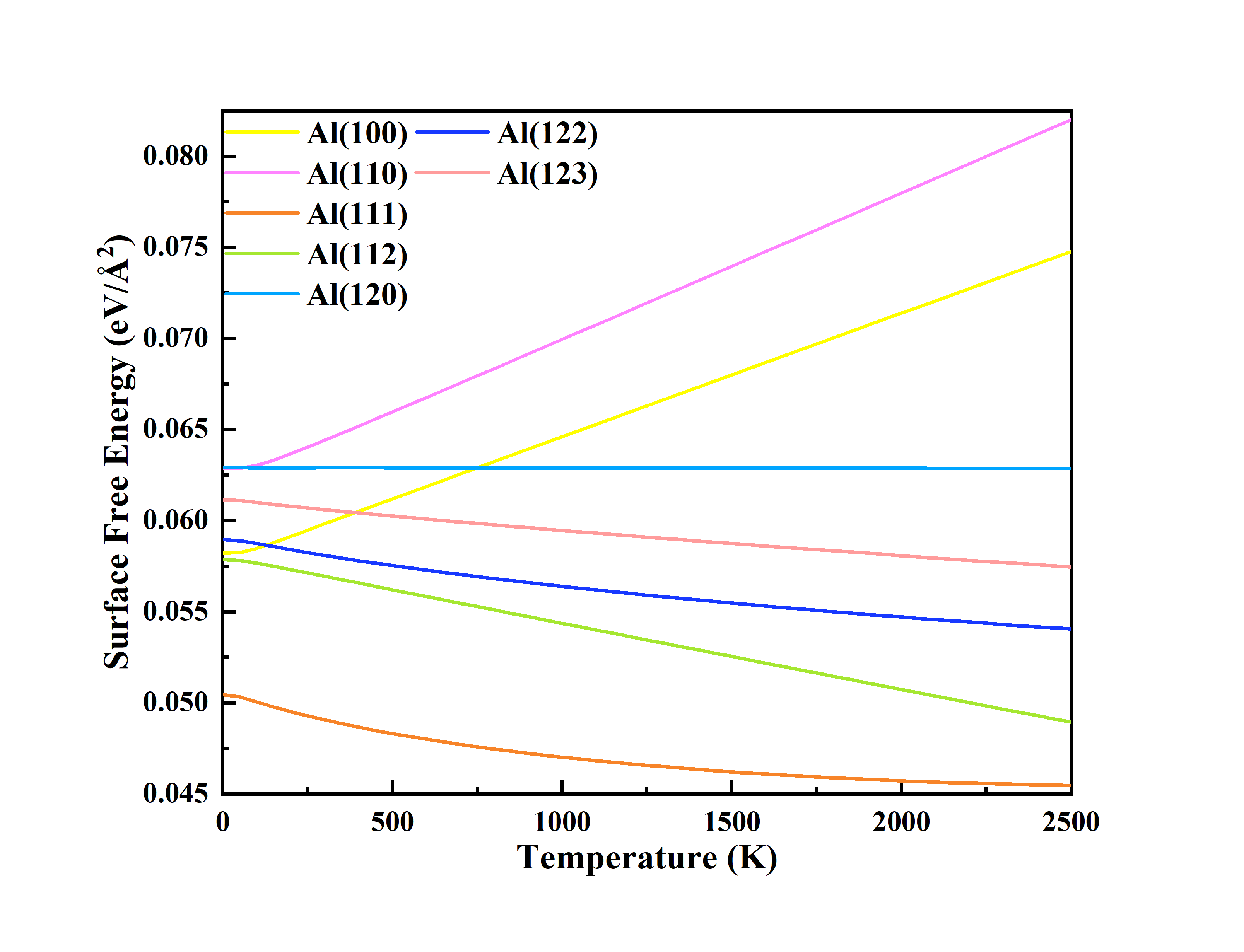}
  \caption{Al surface energy of different planes with temperature considering phonon Gibbs free energy}
  \label{fig:AlSurfE}
\end{figure}

\begin{figure}[htbp]
  \centering
  \includegraphics[scale=0.175]{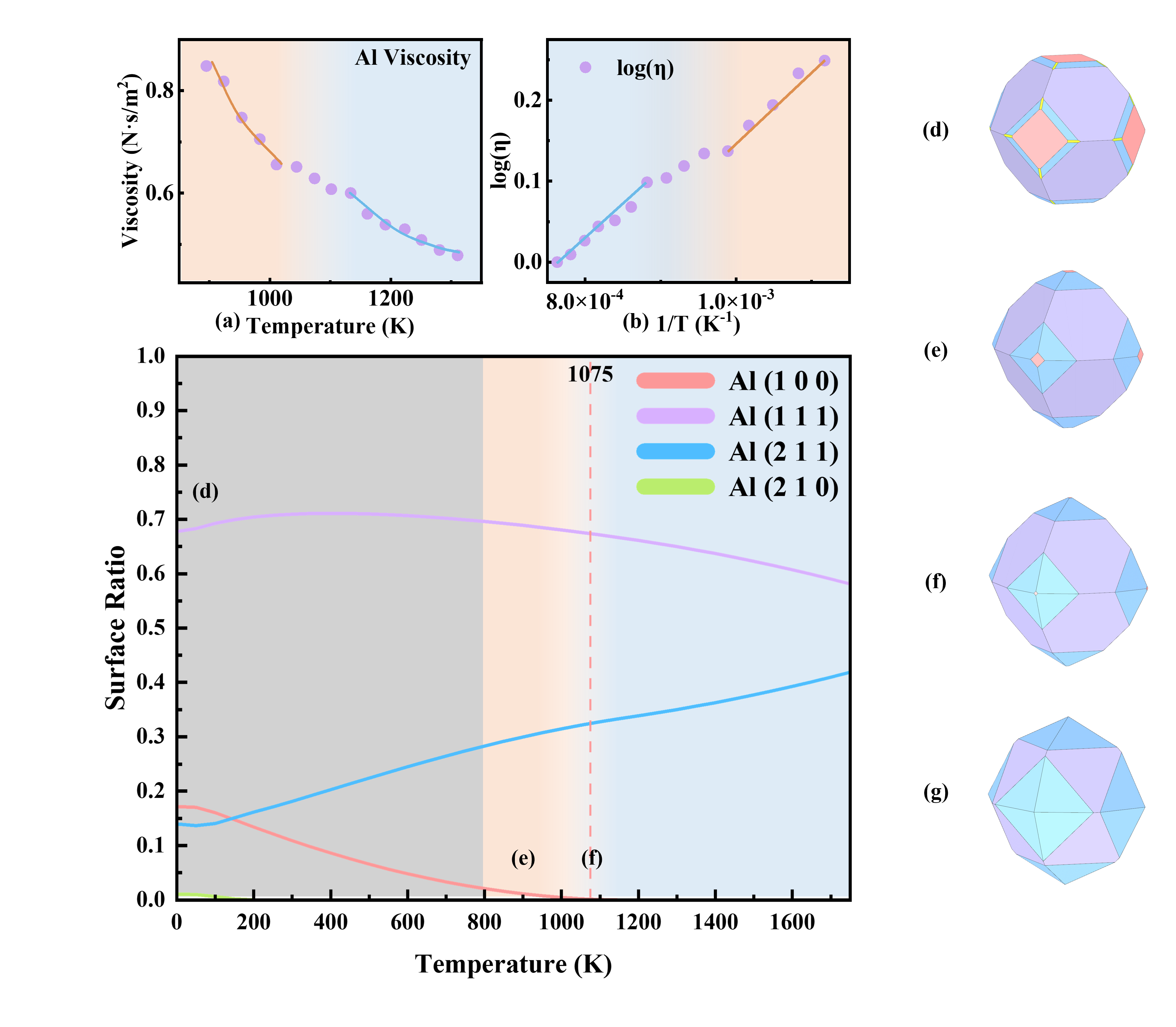}
  \caption{(a)Viscosity-temperature curve of Al melts, (b)viscosity logarithm-temperature reciprocal curve of Al melts\cite{AlVis01}, (c)surface ratio-temperature curve of Al Wulff shape, 
  corresponding Wulff shape are demonstrated in (d)0K (e)900K (f)1100k (g)2500K.}
  \label{fig:AlPho}
\end{figure}

For metal Al, the behavior of Al surfaces with temperature is quite different from that of Pb. As shown in fig.\ref{fig:AlSurfE}, 
the surface energy of Al(100) and Al(110) surfaces increases significantly with temperature (especially, Pb(100) is a surface whose surface energy decreases 
significantly in Pb), which also makes the evolution law of Wulff shape of Al and Pb different with the temperature rising. 
Al(210) surface disappears at 130K, which is of course at solid state. Interestingly enough, 
the surface of Al(100) disappears at 1075k(fig.\ref{fig:AlPho}c), which is coincidentally inside the temperature range of abnormal viscosity breaking point (fig.\ref{fig:AlPho}a\&b).
In fig.\ref{fig:AlPho}a, There is an obvious viscosity rise around 1075K , as well as viscosity logarithm-temperature reciprocal curve shown in fig.\ref{fig:AlPho}b.
The Wulff shapes display clearly that the surface area of Al(100) decreases until it disappears. The surface area of Al(211) increases while that of Al(111) decreases. 
Even so, Al(111) is the surface exposed most in the Wulff shape from first to last.
And at a quite high temperature(fig.\ref{fig:AlPho}g,2500K), the Wulff shape seems to turn into tetrahexahedron from octahedron.

Compared with the results of pure DFT calculation at 0k, its Wulff shape 
exposes more Al(211) surfaces. 
Note that, this temperature is also inside the temperature range of viscosity changes(marked with a fuzzy boundry between orange and blue background colors in Figure.\ref{fig:AlPho}(a)(b)(c))
At the same time, it is easy to find that the calculated temperature point is also very close to the temperature point at which the 
viscosity changes.
And at a quite high temperature(fig.\ref{fig:AlPho}(g),2500K), the Wulff shape seems to turn into tetrahexahedron from octahedron. Surface energy-temperature curve also indicates this.

For both Pb and Al, the structural critical temperature and viscosity leaping temperature show high uniformity, which indicates the close correlation between the structure of melts and its viscosity.
This can also explain the reason that the abnormal viscosity change happens in a relative wide temperature range but not in a certain temperature point. Although in our simulated results, 
the surfaces disappear in a certain temperature, one should keep in mind that our model is only the most probable structure of the melts’ structure distribution. Widening of structure distribution 
and energy fluctuation always exist in the metal melts. 
If the abnormal viscosity change is caused by the change of melts’ structure, another question is how to explain two kinds of viscosity change 
(abnormal viscosity drop (Pb) and viscosity rise (Al) with increasing temperature).

\subsection{Discussion on correlation between viscosity and Wulff shape} \label{sec:wulffVis}

\begin{figure}[htbp]
  \centering
  \subfigure[(100)]{\includegraphics[scale=0.1]{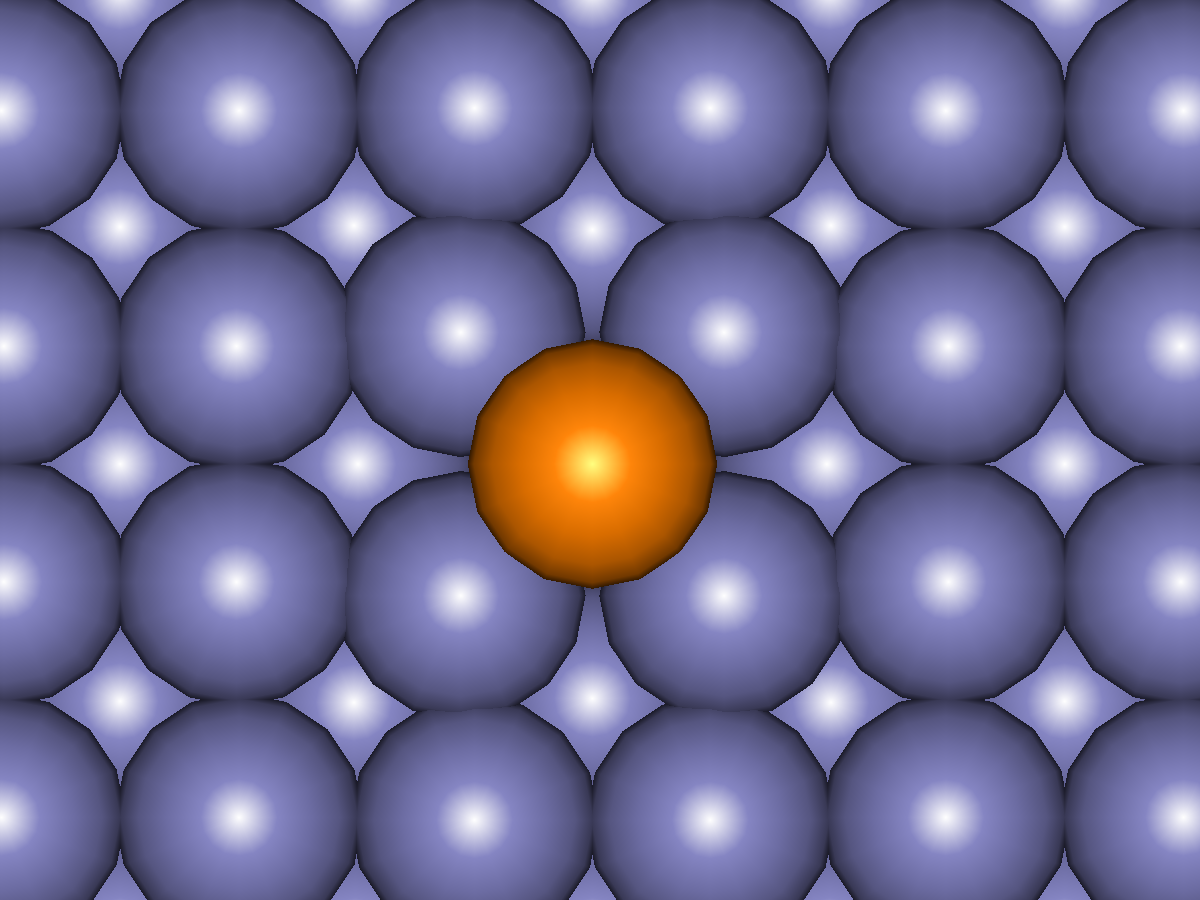} \label{subfig:sur100}} \qquad
  \subfigure[(110)]{\includegraphics[scale=0.1]{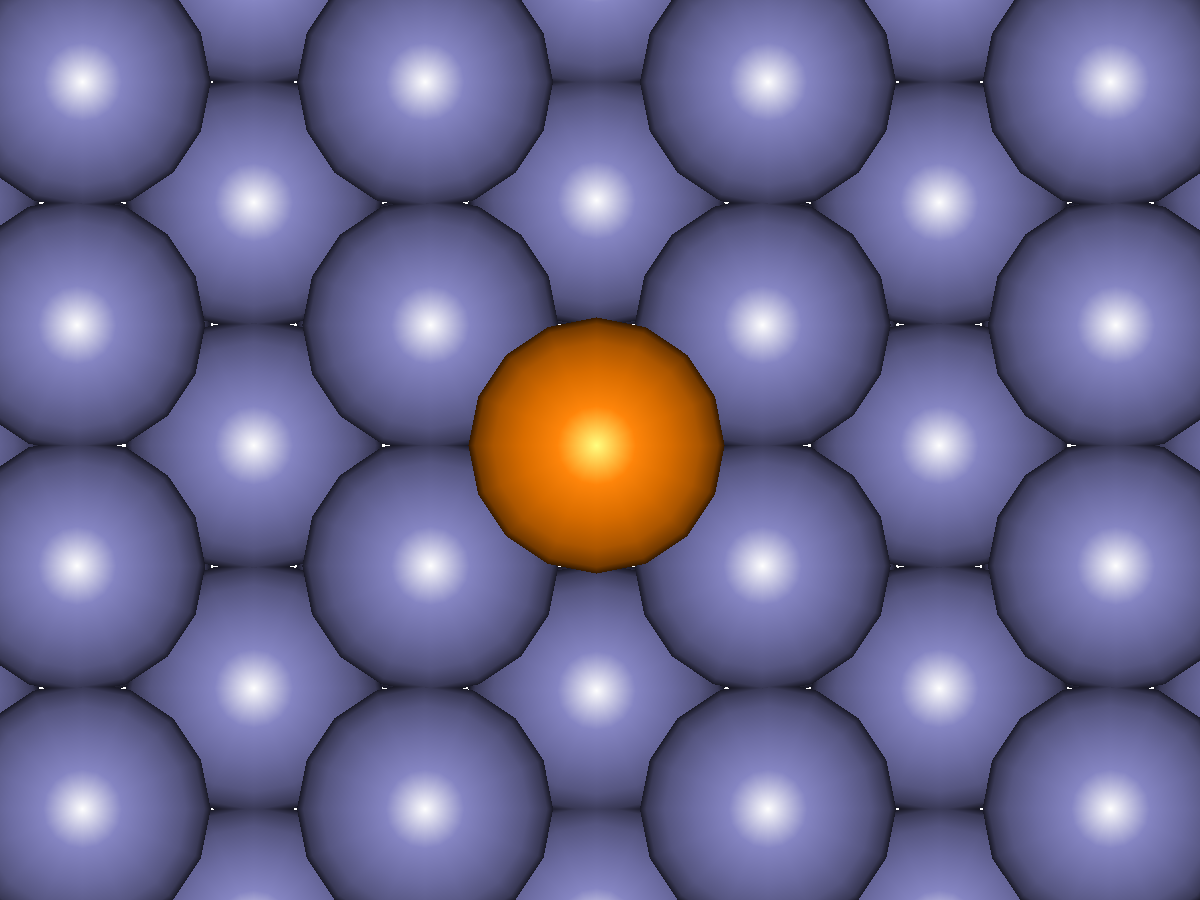} \label{subfig:sur110}} \qquad 
  \subfigure[(111)]{\includegraphics[scale=0.1]{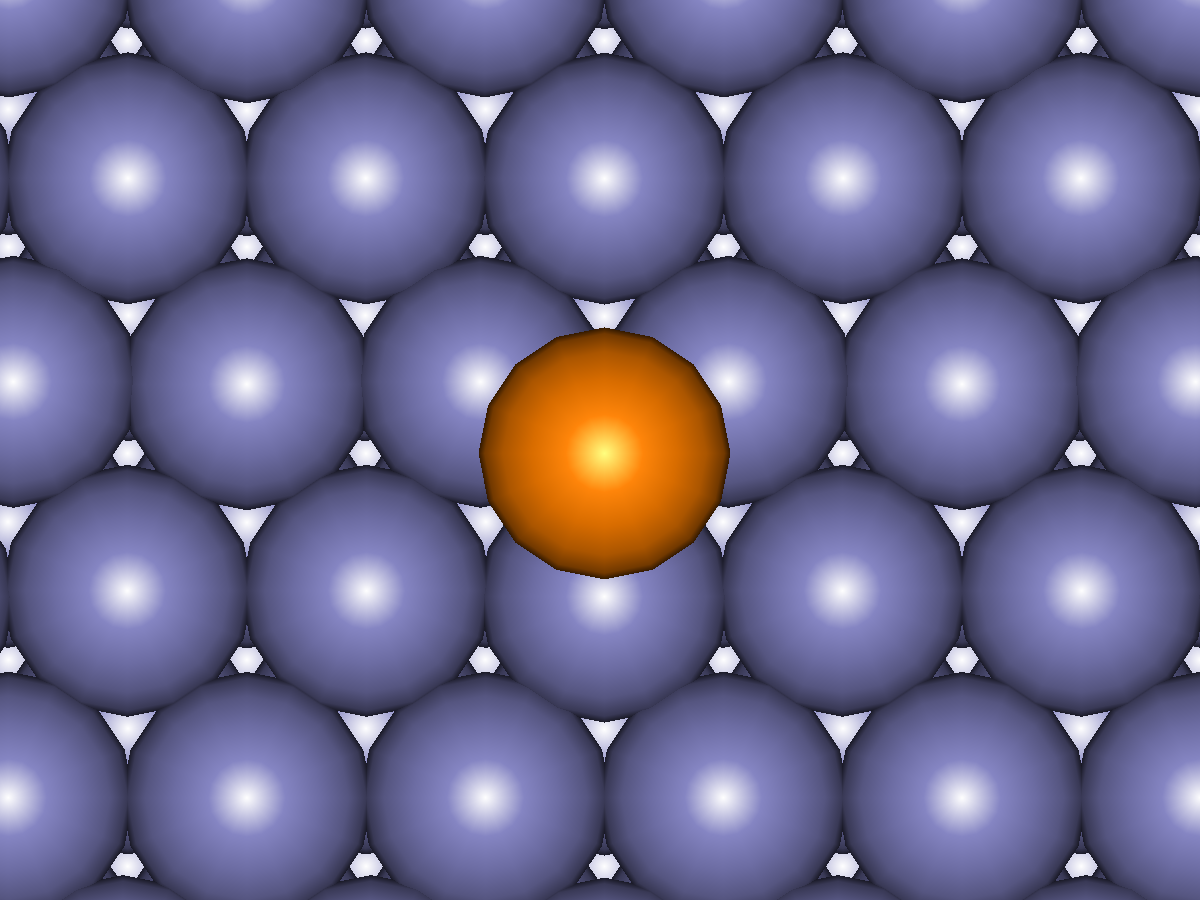} \label{subfig:sur111}} \qquad 
  \subfigure[(112)]{\includegraphics[scale=0.1]{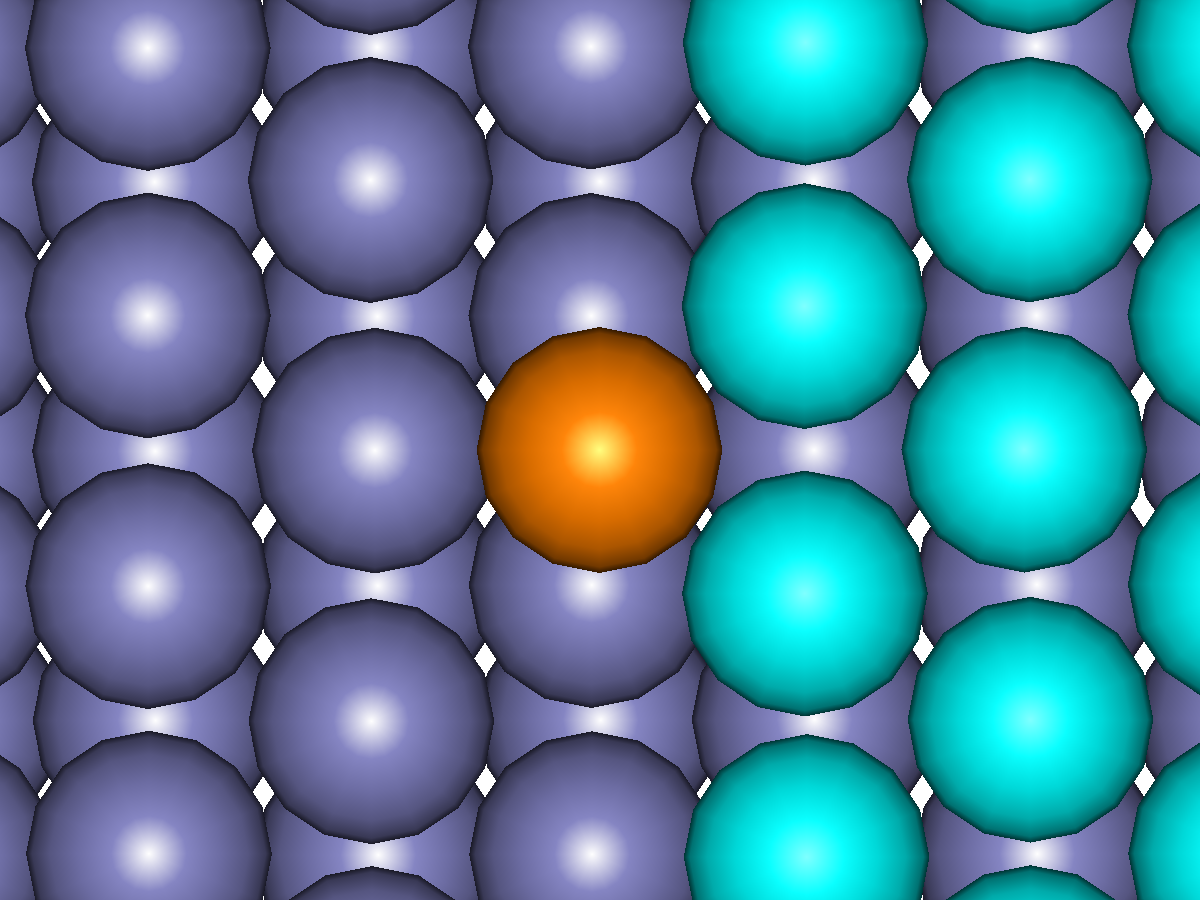} \label{subfig:sur112}} \qquad \\
  \subfigure[(122)]{\includegraphics[scale=0.1]{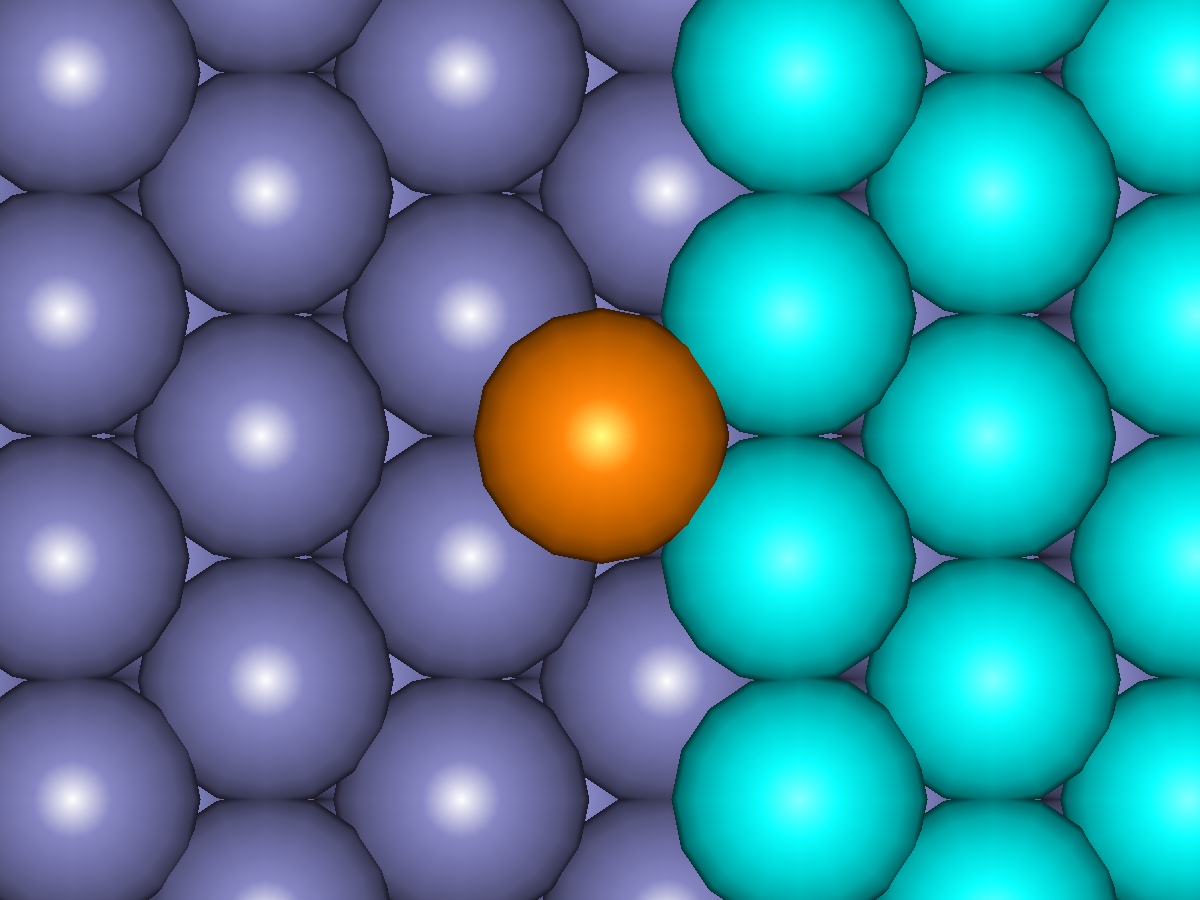} \label{subfig:sur122}} \qquad 
  \subfigure[(210)]{\includegraphics[scale=0.1]{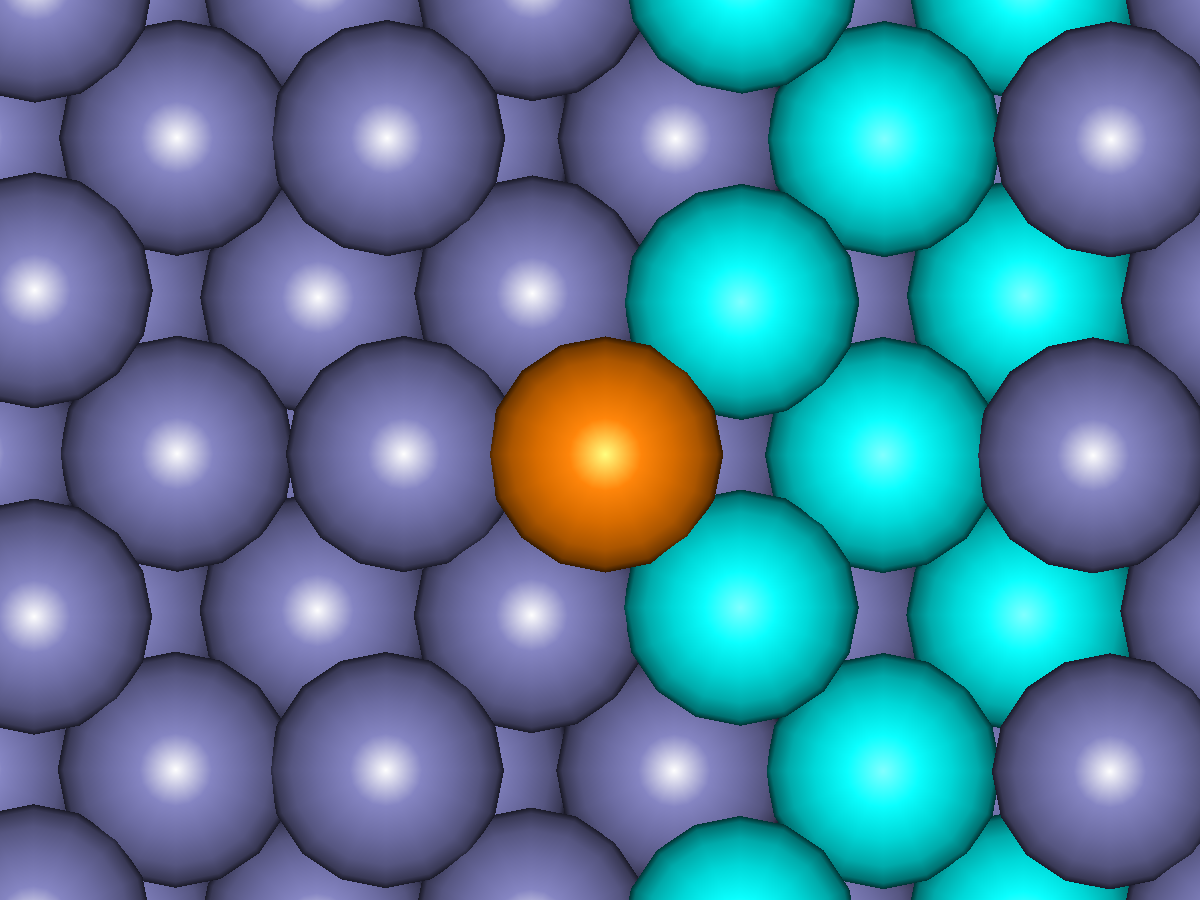} \label{subfig:sur210}} \qquad 
  \subfigure[(321)]{\includegraphics[scale=0.1]{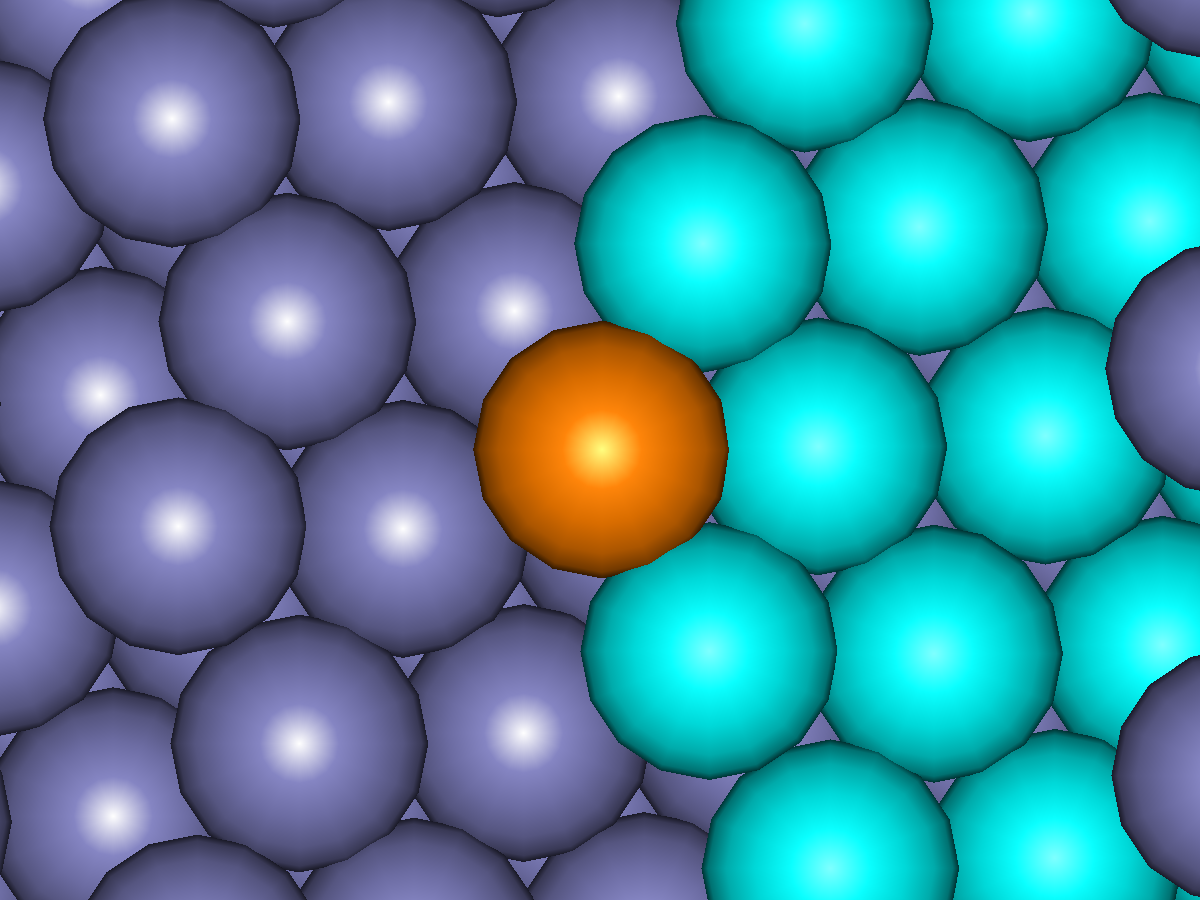} \label{subfig:sur321}} \qquad 
  \caption{Most stable adsorbtion positions in FCC Pb/Al structures,\subref{subfig:sur100} (100) surface \subref{subfig:sur110} (110) surface \subref{subfig:sur111} (111) surface \subref{subfig:sur112} 
  (112) surface  \subref{subfig:sur122} (122) surface \subref{subfig:sur210} (210) surface \subref{subfig:sur321} (321) surface}
  \label{fig:surabs}
\end{figure}

To further verify this conjecture and explain the reason for this coincidence from the atomic level, 
we need a reasonable model to build the correlation between Wulff shape and viscosity. 
Here viscosity was defined as macroscopic effect of average interaction strength of clusters and free atoms,
in other words, it should relate to the interaction strength between parts inside the metallic melts. 
Recall that it has been verified that the metallic melts is composed of clusters and free atoms\cite{FES01,FES02,FES03,FES04}.
In this case, it is reasonable to assume that the viscosity of metals is positively related to the interaction strength of Wulff cluster and free atoms.
To make a easy start, the adsorption energy of a single atom (Pb atom on Pb surfaces/ Al atom on Al surfaces) is used to describe the interaction strength approximatively (shown in fig.\ref{fig:surabs}).
All possible adsorption positions on Pb/Al surfaces exposed in the Wulff shape have been calculated. As expected, the most stable adsorption positions of a single atom on Pb and Al are exactly the same.
For (100) and (110) surfaces are the fourfold hollow site, and for (111) is the threefold hollow site which conforms to the crystal orientation(ABCABC... arrangement in FCC(111)), 
that is, the position that the atom should have occupied in the perfect crystal.
For the (113), (122) and (210) planes, the surface atoms show a stepped arrangement, the most stable adsorption position is 
on the lower side of the step edge, attaching in the mode of expanding adjacent step layers(marked as cyan colored atom layer), which also conforms to the crystal orientation.
Adsorbtion energy calculated with those structures are shown in the Table \ref{tab:DetEne}

\begin{table}[htbp]
  \centering
  \caption{Adsorbtion energy of different surfaces in Wulff shapes of Al and Pb}
  \label{tab:DetEne}
  \begin{tabular}{cccc}
     \toprule
     Type & \makecell{Adsorbtion Energy(per atom)\\$(\rm eV)$} & Type & \makecell{Adsorbtion Energy(per atom)\\$(\rm eV)$}  \\
     \midrule
     ${\rm Al(100)}$ & $-2.98$ & ${\rm Pb(100)}$ & $-2.95$\\
     ${\rm Al(110)}$ & $-3.77$ & ${\rm Pb(110)}$ & $-3.41$\\
     ${\rm Al(111)}$ & $-3.00$ & ${\rm Pb(111)}$ & $-2.85$\\
     ${\rm Al(210)}$ & $-3.75$ & ${\rm Pb(321)}$ & $-3.51$\\
     ${\rm Al(211)}$ & $-3.54$ & ${\rm Pb(211)}$ & $-3.51$\\
     ${\rm Al(221)}$ & $-3.70$ & ${\rm Pb(221)}$ & $-3.37$\\
     \bottomrule
  \end{tabular}
\end{table}

As shown in Table \ref{tab:DetEne}, 
the adsorption energy of a single atom on Pb(100) and Al(100) surface is relatively low,
and adsorption energy of Pb(321) surface is highest in all Pb surfaces, similar with that on Al(311) surface. 
Recall that the Pb(321) surface that has the highest interaction strength disappears at 975K, while the abnormal viscosity drop happens at the same temperature range. 
In other words, the part with strong cluster-liquid interaction disappear, which will make the average interaction relatively weak (viscosity drop). 
Oppositely, Al(100) surface that has the lowest interaction strength disappears, at the same temperature range around 1075K, 
the abnormal viscosity rise has been observed. All evidence indicates that the uniformity between the abnormal viscosity change and the change of Wulff shape is not a coincidence. 
The replacement of some surfaces of statistical Wulff shape is expected to be reflected in the macro properties of the melt (viscosity).
The abnormal  viscosity change is caused by the significant structure change of cluster (short range ordering) in the metallic melts.
In fact, the cluster structure-temperature relationship is mathematically continuous, But the trend of structure change is totally different before and after the break point. 
To further understand the micro-mechanism of viscosity under atomic level, 
the quantitative relationship between viscosity and the melts’ structure should be determined, which would be our research objective of the next stage.

\section{Conclusion}

In this paper, the Wulff cluster  model combined phonon calculation is used to investigate the relationship between the structure of metallic melts and the abnormal viscosity change. 
Although absolute value of the surface energy does not change significantly with temperature, the Wulff shape changed evidently. At about 975K, Pb(321) surface completely disappears, 
while the abnormal viscosity drop happens at the same temperature range.  Oppositely, Al(100) surface disappears, at the same temperature range around 1075K, the abnormal viscosity 
rise has been observed. Pb(321) surface has the highest interaction strength, while Al(100) has the lowest. The abnormal viscosity drop corresponds to the disappearance of surface 
with the highest interaction strength (Pb(321) surface), while the rise one corresponds to the disappearance of surface with the lowest adsorption energy (Al(100) surface). 
All evidence indicates that the uniformity between the abnormal viscosity change and the change of Wulff shape is not a coincidence. The replacement of some surfaces of 
statistical Wulff shape is expected to be reflected in the macro properties of the melt (viscosity). 
A possible mechanism of the abnormal viscosity change is that it caused by the significant structure change of cluster (short range ordering) in the metallic melts.

\section{Acknowledgment}
The authors gratefully acknowledge the financial support from the Basic reasearch priorities program of national natural science foundition of Shandong, China (Project No. ZR2021ZD22),
the China Postdoctoral Science Foundation (Project No. 2018M642642) 
and the National Natural Science Foundation of China (Project No. U1806219).

\bibliographystyle{unsrt}
\bibliography{bibs}

\end{document}